\documentclass[pdflatex,sn-mathphys-num,10pt]{sn-jnl}

\usepackage{graphicx}%
\usepackage{multirow}%
\usepackage{amsmath,amssymb,amsfonts}%
\usepackage{amsthm}%
\usepackage{mathrsfs}%
\usepackage[title]{appendix}%
\usepackage{xcolor}%
\usepackage{manyfoot}%
\usepackage{booktabs}%
\usepackage{algorithm}%
\usepackage{algorithmicx}%
\usepackage{algpseudocode}%
\usepackage{listings}%
\usepackage{bbm}
\usepackage{cleveref}

\usepackage{stix} 

\theoremstyle{thmstyleone}%
\theoremstyle{thmstyletwo}%
\newtheorem{remark}{Remark}%

\theoremstyle{thmstylethree}%

\graphicspath{{figs/}}

\newcommand{\body}{\mathcal{B}}
\newcommand{\Mcal}{\mathcal{M}}

\newcommand{\added}[1]{{#1}}

\usepackage[normalem]{ulem}

\begin{document}

\title[Creep and rough surfaces : evolution of the contact between rough viscoelastic solids]{Evolution of the contact between rough viscoelastic solids after decreasing loads: memory erasure and monotonic increase}


\author[1]{\fnm{Zichen} \sur{Li}}\email{zichen.li@sorbonne-universite.fr}

\author[1]{\fnm{Renald} \sur{Brenner}}\email{renald.brenner@sorbonne-universite.fr}

\author*[1]{\fnm{Lucas} \sur{Frérot}}\email{lucas.frerot@sorbonne-universite.fr}

\affil[1]{Sorbonne Université, CNRS, Institut Jean Le Rond d’Alembert, F-75005 Paris, France}


\abstract{

The real area of contact governs, in part, the magnitude of the friction force, yet its time evolution in rough viscoelastic interfaces remains incompletely understood. In experiments of contact between polymethylmethacrylate blocks under decreasing normal loads, Dillavou and Rubinstein have shown that the true contact area exhibits, after unloading, a decreasing phase and long-term memory of the contact state prior to unloading. It is however unclear what modeling ingredients are necessary to reproduce these two features. Here, we investigate these effects using fractional viscoelastic rough contact models. By adapting existing contact theories and numerical simulation methods to fractional viscoelasticity, which induces a wide relaxation spectrum, we reproduce logarithmic aging under constant load, but show that memory of the contact state is erased upon unloading. Indeed, the contact area behaves as if it had always experienced the reduced load, even on short time-scales, contrasting with the response of a standard linear solid. Moreover, none of our results show a decreasing regime of the contact area after unload: we ultimately prove that this is the case for all linear viscoelastic models---despite capturing logarithmic aging---leading to the conclusion that additional local internal variables are required to explain both long-term contact memory and contact area reduction after unloading.

}




\maketitle

\section{Introduction}\label{sec1}

Understanding friction and wear is not merely academic—it has quantifiable environmental and economic stakes. \Citet{jostTribologyMicroMacro2005} mentions that the application of tribological principles and practices can lead to savings of 1.0\% to 1.4\% of an industrial country's GNP. However, efficient improvements in the design of sliding interfaces require understanding of friction, particularly in transient conditions.

\begin{figure}
    \centering\includegraphics{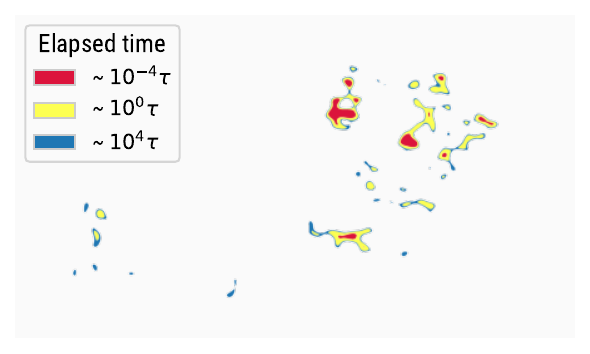}
    \caption{Illustration of the evolution of the true contact area in a rough contact with fractional viscoelastic materials under constant load. $\tau$ is a characteristic time.}\label{fig:area_evolution}
\end{figure}

At rest, the force required to macroscopically slide two rough surfaces in contact increases logarithmically over time~\citep{coulombTheorieMachinesSimples1821,dieterichTimedependentFrictionRocks1972}. Depending on the system, this growth can be attributed to a combination of creep~\citep{dieterichDirectObservationFrictional1994}, which increases the true contact area, and chemical~\citep{liChemicalAgingLargescale2018} or structural aging~\citep{frerotMolecularMultiasperityContacts2023}, which both increase the shear resistance of micro-contacts. Unfortunately, these mechanisms cannot be distinguished macroscopically under monotonic loading, as they all similarly contribute to the increase of the friction force. However, all of them are mediated through the true contact area: we therefore focus on its evolution due to viscoelasticity, as illustrated in \cref{fig:area_evolution} under monotonic loading.

Recent experiments by \citet{dillavouNonmonotonicAgingMemory2018} have shown that, when the normal force bringing two polymethylmethacrylate (PMMA) blocks in contact is reduced after a rest period, the true contact area \emph{decreases} over a time comparable to the rest time, highlighting long-term memory effects. This non-monotonic behavior cannot be explained within the classical framework of rate-and-state friction laws, which postulate a monotonic evolution of the contact area if the load is held constant~\citep{baumbergerSolidFrictionStick2006}.

In this article, we investigate with analytical models and numerical simulations the behavior of rough viscoelastic contacts under non-monotonic loading conditions. We provide a systematic way of modeling viscoelastic contacts with a controlled relaxation spectrum breadth through the use of fractional viscoelastic models. This allows us to probe the long-term memory effects observed in experiments and the reduction of the true contact area after unloading.

We first present the analytical and numerical methods used to model viscoelastic rough contacts with fractional viscoelasticity constitutive models. We then discuss results under constant loading conditions, where we can reproduce the experimentally observed logarithmic aging, and under decreasing loading conditions, where we show that long-term memory effects are absent, and discuss the reasons and possible modeling paths to remediate this absence.

\section{Methods}\label{sec2}
We consider the contact between a rigid rough surface and a flat, deformable half-space $\body$, whose behavior is linear, non-aging viscoelastic with \emph{finite} instantaneous and long-time elastic responses respectively characterized by Young's moduli $E_0$ and $E_\infty$. In one dimension, under a stress loading history $\sigma(t)$, the strain $\varepsilon(t)$ can thus be expressed with the following Stieltjes integral:
\[ \varepsilon(t) =  \frac{1}{E_\infty} \int_{-\infty}^tJ(t - \tau)\frac{\mathrm d\sigma}{\mathrm dt}(\tau)\,\mathrm d\tau, \]
where $J$ is the non-dimensional creep function, $E_\infty$ is the long-time elastic response of the material, and the derivative operation $\mathrm d/\mathrm dt$ is understood in the sense of distributions, i.e., with appropriate Dirac distributions where $\sigma$ is discontinuous. Conversely, when the strain history is prescribed, the stress can be obtained with a similar integral:
\[ \sigma(t) = E_\infty \int_{-\infty}^t G(t - \tau)\frac{\mathrm d\varepsilon}{\mathrm dt}(\tau)\,\mathrm d\tau, \]
where $G$ is the non-dimensional relaxation function. The instantaneous and long-time values of the creep and relaxation functions are thus
\[
J(+\infty) = G(+\infty) = 1\quad\text{and} \quad J(0) = \frac{1}{G(0)}=k\quad\text{with}\quad k = \frac{E_\infty}{E_0}<1
\]
\begin{remark}\label{rem:transform}
  $J$ and $G$ are non-negative and monotonic, with $J$ increasing and $G$ decreasing. The two integrals defined above through $J$ and $G$ can be thought of as a transform pair, i.e., if two functions $\sigma$ and $\varepsilon$ satisfy one relation, they also satisfy the other.
\end{remark}

\subsection{Contact models}
We first describe the analytical approach we use to solve the rough contact problem, which follows the procedure first laid out by \citet{10.1115/1.3644020}, who established the equations governing the evolution of the contact area while it increases, followed by \citet{tingContactStressesRigid1966}, who resolved the issue of a decreasing contact area. The Hertzian case is nicely summarized by \citet{johnsonContactMechanics1985}, but here we apply this procedure to rough contacts, using Persson's model~\citep{perssonTheoryRubberFriction2001}, as was first done by \citet{papangeloViscoelasticNormalIndentation2021}. Without loss of generality, we restrict our study to the small contact regime, such that the true contact area $\bar A$ can be expressed with the linearised expression:
\begin{equation}
  \label{eq:area_with_dims}\frac{\bar A(t)}{A_0} = \frac{2}{h'_\mathrm{rms} E^*_\infty} \int_{-\infty}^t J(t - \tau)\frac{\mathrm d \bar p}{\mathrm dt}(\tau)\,\mathrm d\tau,
\end{equation}
where $A_0$ is the nominal contact area, $h'_\mathrm{rms}$ is the root-mean-square of the slopes of the rough surface, $E^*_\infty = E_\infty/(1-v^2)$ ($v$ is the Poisson coefficient) and $\bar p(t)$ is the applied normal pressure. To simplify expressions, we hereafter define $A = \bar A/A_0$ as the normalized contact area and $p = 2\bar p / (h'_\mathrm{rms} E^*_\infty)$ as the normalized pressure, such that \cref{eq:area_with_dims} can be written as:
\begin{equation}
  \label{eq:area} A(t) = \int_{-\infty}^t J(t - \tau) \frac{\mathrm dp}{\mathrm dt}(\tau)\,\mathrm d\tau.
\end{equation}
Note that \cref{eq:area} is only valid as long as the area is increasing, i.e. $t \leq t_m$, where $A(t_m)$ is maximum. When the contact area decreases ($t \geq t_m$), one needs to invert the relationship between $A$ and $p$ using the integral defined with the relaxation function $G$ (cf. \cref{rem:transform}), and solve for a time $t_1 \leq t_m$ such that:
\begin{equation}
  \label{eq:t1} p(t) = \int_{-\infty}^{t_1}G(t - \tau)\frac{\mathrm dA}{\mathrm dt}(\tau)\,\mathrm d\tau.
\end{equation}
The time $t_1$ is the time during the increasing phase of $A$ when $A(t_1) = A(t)$, with $A(t_1)$ computed using \cref{eq:area}. While \cref{eq:t1} is in general not tractable analytically, it lends itself well to non-linear root finding algorithms such as Newton's method.

\subsection{Viscoelastic behavior}\label{viscoelastic behavior}
In this section, we present the viscoelastic constitutive models used in both analytical models and simulations.

\subsubsection{Fractional viscoelasticity}

Most experiments investigating frictional aging highlight a logarithmic dependence of the friction force on the contact time. In a viscoelastic material, this dependency is in part explained by the logarithmic increase of the true contact area under constant load. Logarithmic creep behavior, typical of disordered systems, is the outcome of a rich relaxation spectrum~\citep{amirRelaxationsAgingVarious2012}, with characteristic times distributed as a power-law. To reproduce this effect, we investigate the contact properties of materials which follow a fractional viscoelastic law, more specifically the fractional Zener model, similar to the Standard Linear Solid (SLS) model, but with the standard dashpot replaced with a Scott--Blair element~\citep{blairEstimationFirmnessSoft1943}, see \cref{fig:comparasion}b. The Scott--Blair element obeys the following relationship for its stress $\sigma_\mathrm{SB}$ as function of strain:
\[ \sigma_\mathrm{SB} = E' \frac{\mathrm d^\nu \varepsilon}{\mathrm dt^\nu}, \]
where $\nu \in [0, 1]$ is the order of the Caputo fractional derivative $\mathrm d^\nu / \mathrm dt^\nu$ \citep{caputoLinearModelsDissipation1967,caputoNewDissipationModel1971}, and $E'$ is a constant homogeneous to $\mathrm{pressure}\cdot\mathrm{time}^\nu$. While the details of such model go beyond the scope of this article\footnote{The interested reader is referred to the review of \citet{mainardiCreepRelaxationViscosity2011, mainardiFractionalCalculusWaves2022} for a didactic introduction.}, we want to point out that $\nu$ smoothly interpolates between a fully elastic behavior ($\nu = 0$) and a standard linear solid ($\nu = 1$), with intermediate values giving rise to wide relaxation spectra and long-term memory \citep{lakesViscoelasticMaterials2009,bagleyTheoreticalBasisApplication1983}. As demonstrated in \citep{torvikAppearanceFractionalDerivative1984}, one or two fractional derivative terms in the constitutive equation are sufficient to represent the response of actual materials over a wide range of time scales. Unlike a multi-branch generalized Maxwell model, only one parameter, $\nu$, controls the breadth of the relaxation spectrum, and it can be measured experimentally, such as with a stress relaxation test~\citep{hernandez-jimenezRelaxationModulusPMMA2002,huNoteFractionalMaxwell2011}.
The creep and relaxation functions of the fractional Zener model are given by~\citep{mainardiCreepRelaxationViscosity2011}:
\begin{equation*}
  \begin{aligned}
    J(\bar t)  & = 1 - (1 - k)E_\nu(-(\bar{t}/\tau_C)^\nu)\\
    G(\bar t)  & = 1 + \frac{1 - k}{k}E_\nu(-(\bar t / \tau_R)^\nu)
  \end{aligned}\quad \text{with }E_\nu(\xi) = \sum_{n = 0}^\infty \frac{\xi^n}{\Gamma(1 + \nu\cdot n)},
\end{equation*}
where $\tau_C$ and $\tau_R = k^\frac{1}{\nu}\tau_C$ are the creep and relaxation timescales respectively, $E_\nu$ is the Mittag--Lefler function, and $\Gamma$ is the Gamma function (which generalizes the factorial to real numbers). For simplicity, we introduce the non-dimensional time $t = \bar t / \tau_C$, such that:
\begin{equation}
    \label{eq:relax_functions}
\begin{aligned}
    J(t) & = 1 - (1 - k)E_\nu(-t^\nu),\\
  G(t) & = 1 + \frac{1 - k}{k}E_\nu(-t^\nu/k).
\end{aligned}
\end{equation}
Thus, fractional viscoelastic models only effectively have two parameters, $k$ and $\nu$.

\begin{remark}
  The Mittag--Leffler function generalizes the exponential, and is the solution to the fractional differential equation $\frac{\mathrm d^\nu y(t)}{\mathrm dt^\nu} = y(t)$, with $y(0) = 1$. We then have $E_1(t) = e^t$.
\end{remark}

\begin{remark}
  The time derivative of $E_\nu(-t^\nu)$ is expressed as~\citep{mainardiCreepRelaxationViscosity2011}:

  \[ \frac{\mathrm d }{\mathrm dt}E_\nu(-t^\nu) = -t^{\nu-1} E_{\nu,\nu}(-t^\nu),\quad\text{with}\quad E_{\nu,\beta}(\xi) = \sum_{n = 0}^\infty \frac{\xi^n}{\Gamma(\beta + \nu\cdot n)}. \]
\end{remark}

\subsubsection{Standard Linear Solid model}

The Standard Linear Solid (SLS), also known as the Zener model, can be represented by a spring in parallel with a linear dashpot, and the pair in series with a second spring.
In non-dimensional form, the creep and relaxation functions of the standard linear solid model are given by:
\begin{equation*}
    \begin{aligned}
        J(t) & = 1-(1-k)e^{-t}, \\
        G(t) & = 1+\frac{1-k}{k}e^{-t/k},
    \end{aligned}    
\end{equation*}
\added{which can be seen as the special case of \cref{eq:relax_functions} when $\nu = 1$.}

\subsubsection{Generalized Maxwell model}

\begin{figure}
\centering
\includegraphics[width=1\linewidth]{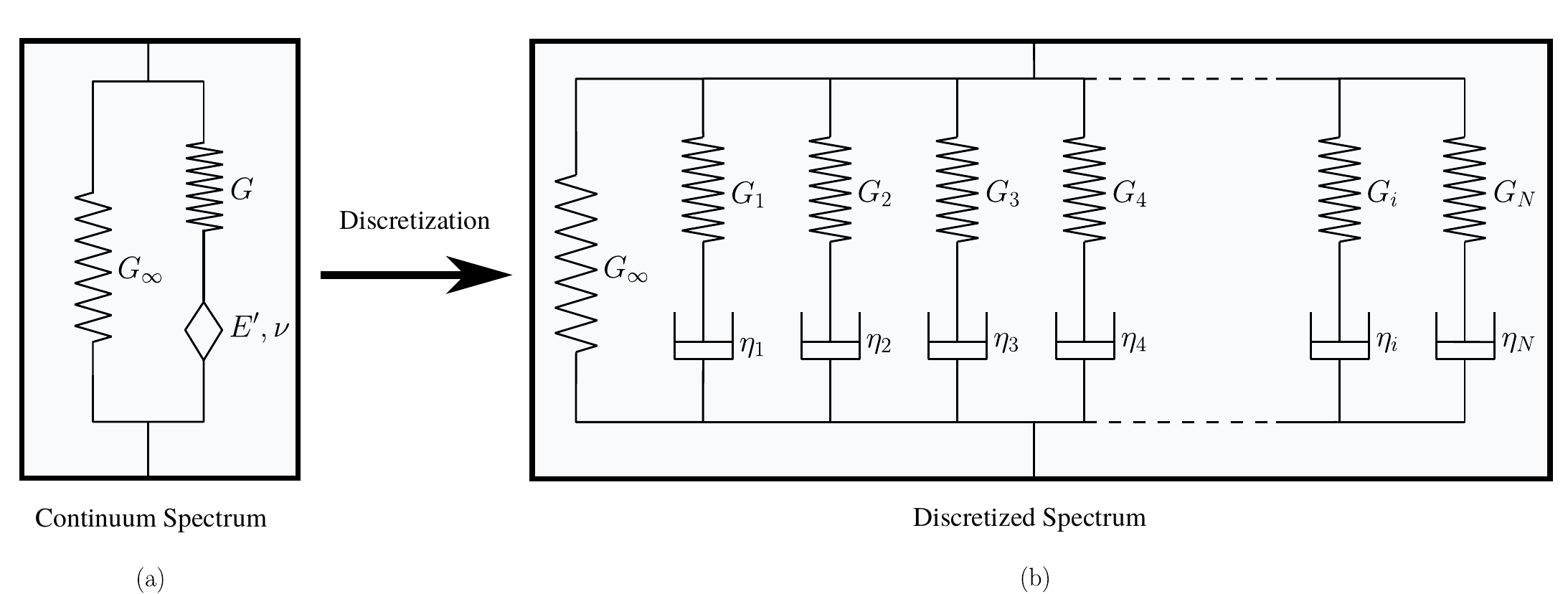}
\caption{Discretization of linear viscoelastic model (a) Fractional Zener model with continuum spectrum (b) Generalized Maxwell model with discretized spectrum: an elastic branch with shear modulus ($G_\infty$) in parallel with multiple Maxwell branches composed of a spring ($G_i$) and a dashpot ($\eta_i$). }
\label{fig:comparasion}
\end{figure}

The generalized Maxwell model, also known as the Wiechert model, generalizes the Standard Linear Solid by allowing any number of relaxation times, see~\cref{fig:comparasion}. While being versatile, it has the drawback of having many adjustable independent variables (as many as there are relaxation times), and thus requires specialized methods to fit these parameters to experiments~\citep{jalochaRevisitingIdentificationGeneralized2015}. However, the discrete nature of its relaxation spectrum makes it easy to use in simulations, as we now demonstrate.

\subsection{Numerical simulations}
While the analytical contact model above gives an accurate evolution of the true contact area, it only gives information on the area and hides the other relevant quantities of the contact, such as the pressure distribution. We give here a robust procedure for the simulation of rough, fractional viscoelastic contacts.

\subsubsection{Discrete spectrum}
Unfortunately, while the fractional Zener model allows explicit analytical expressions of the contact area (since both $J$ and $G$ are known analytically), the Mittag--Leffler function is not suitable for existing simulation methods of viscoelastic contacts, which assume the discrete spectrum of a generalized Maxwell (or similar) model, i.e. a sum of parallel Maxwell elements (spring and dashpot in series, see \cref{fig:comparasion}).

To resolve this issue, we improved the spectrum discretization procedure proposed by \citet{vernierHomogenizationCompositeMaterials2023} \added{and \citet{papouliaRheologicalRepresentationFractional2010a}}, and express the relaxation function $G(t)$ as a Prony series:
\begin{equation}
  \label{eq:prony} G(t) = 1 + \frac{1 - k}{k}E_\nu(-t^\nu / k) \approx 1 + \sum_{n=1}^{N_e} G_n e^{-t/\tau_n}
\end{equation}
where $\{G_n\}$ (set of non-dimensional elastic coefficients) and $\{\tau_n\}$ (set of non-dimensional relaxation times) are determined by a quadrature approximation of the following integral representation of $E_\nu(-t^\nu)$ \citep{gorenfloComputationMittagLefflerFunction2002}\footnote{The reader should be aware that some references, such as \citep{gorenfloComputationMittagLefflerFunction2002}, include a typo in the expression of this integral, in the form of an additional $\xi$ factor.}:
\[ E_\nu(-t^\nu) = \frac{\sin\pi\nu}{\pi}\int_0^\infty \frac{\xi^{\nu-1}}{1 + 2\xi^\nu\cos\pi\nu + \xi^{2\nu}}e^{-\xi t}\,\mathrm d\xi. \]
We refer the reader to appendix \ref{app:ml_discrete} and the supplementary material \citep{liSupplementaryCodesData2025} for the details on the expressions for $G_n$ and $\tau_n$. Once established, the Prony series can readily be combined with existing methods for rough contact mechanics.

\subsubsection{Contact method}
\citet{bugnicourtTransientFrictionlessContact2017} devised a backward Euler time integration scheme for contact problems with a generalized Zener model (i.e., Zener elements in parallel). Here, we adapt their method to a generalized Maxwell model, represented by the Prony series of \cref{eq:prony}.

The surface normal displacement $u_z$ of an elastic half-space, with shear modulus $\mu$, in equilibrium is given by the convolution of the Boussinesq fundamental solution with the normal surface pressure $p_z$:
\begin{equation}
  \label{eq:boussinesq} \mu u_z(x, y) = \int_\body \frac{1-v}{2\pi} \frac{p_z(x', y')}{\sqrt{(x - x')^2 + (y - y')^2}}\,\mathrm dx'\mathrm dy' \quad \Leftrightarrow \quad \mu u_z = \Mcal[p_z].
\end{equation}
Since the viscoelastic contact problem is strictly equivalent to an elastic contact problem in the Laplace domain, the Boussinesq solution can be employed for each branch of the generalized Maxwell model, i.e.
\begin{equation}
  \label{eq:maxwell} \frac{1}{\tau_n}\Mcal[p_{z,n}(t)] + \Mcal\left[\frac{\mathrm dp_{z,n}}{\mathrm dt}\right] = G_n \frac{\mathrm du_z}{\mathrm dt},
\end{equation}
where $p_{z,n}$ is the normal pressure in the $n$-th branch of the model. The elastic branch of the model follows \cref{eq:boussinesq} with $\mu = 1$ (in non-dimensional units), and we note $p_{z,0}$ its pressure. Using a backward Euler scheme for the time derivatives, we obtain the time-discrete version of \cref{eq:maxwell} for time $t + \Delta t$:
\[ \frac{1}{\tau_n}\Mcal[p_{z,n,t + \Delta t}] + \frac{1}{\Delta t}\left(\Mcal[p_{z,n,t +\Delta t}] - \Mcal[p_{z,n, t}]\right) = \frac{G_n}{\Delta t}\left(u_{z,t+\Delta t} - u_{z, t}\right). \]
By summing over all branches of the model (excluding the elastic branch), we obtain:
\[ \Mcal\left[\sum_{n=1}^{N_e} p_{z,n,t+\Delta t}\right] - \sum_{n=1}^{N_e} \gamma_n \Mcal[p_{z,n,t}] + \bar G u_{z,t} = \bar G u_{z,t + \Delta t} \]
with $\gamma_n = \frac{\tau_n}{\Delta t + \tau_n}$ and $\bar G = \sum_{n=1}^{N_e} \gamma_n G_n$. Finally, adding the elastic branch, we obtain:
\begin{equation}
  \label{eq:euler} \Mcal[p_{z, t+\Delta t}] - \sum_{n=1}^{N_e}\gamma_n\Mcal[p_{z,n,t}] + \bar G u_{z,t} = (1 + \bar G)u_{z, t+\Delta t},
\end{equation}
where $p_{z,t + \Delta t} = \sum_{n=0}^{N_e}p_{z,n,t+\Delta t}$ is the total contact pressure, including the elastic branch $n = 0$. One can notice that \cref{eq:euler} is similar to \cref{eq:boussinesq}, only with an offset known \emph{apriori}:
\begin{equation}
  \Delta h_{t} = \bar G u_{z,t} - \sum_{n=1}^{N_e} \gamma_n\Mcal[p_{z,n,t}].
\end{equation}
As pointed out by \citet{bugnicourtTransientFrictionlessContact2017}, this offset can be transferred to the rough surface, allowing the reuse of Polonsky–Keer’s projected conjugate-gradient solver for elastic contact~\citep{polonskyNumericalMethodSolving1999}; or other related bound-constrained CG variants~\citep{vollebregtNewSolverElastic2014}. The algorithm only needs additional storage of the $\Mcal[p_{z,n,t}]$ fields. The detailed contact procedure is outlined in \cref{algorithm}.

\begin{algorithm}
\caption{Viscoelastic contact routine based on generalized Maxwell model}\label{algorithm}
\begin{algorithmic}
\State Input: external load $W_{t+\Delta t}$, timestep $\Delta t$, previous state fields $\mathbb{M}_n$, previous displacement $\mathbb U$, surface $\mathbb H$, Maxwell parameters, $G_n$, $\tau_n$, $k\in[1, N_e]$.

\State $\gamma_n \gets \frac{\tau_n}{\tau_n+\Delta t}$
\State $\bar G \gets \sum_{k=1}^{N_e}\gamma_n G_n$
\State $\Delta \mathbb H\gets \bar G \mathbb U - \sum_{k=1}^{N_e}\gamma_n \mathbb M_n$
\State ${\mathbb H}' \gets (1 + \bar G)\mathbb H - \Delta \mathbb H$
\State Solve an elastic contact problem with $\mathbb H'$ as the rough surface, $W_{t+\Delta t}$ as the normal load, and elastic shear modulus $\mu = 1$. The obtained total contact pressure is $\mathbb P$.
\State $\mathbb U_\mathrm{new} \gets \frac{1}{1 + \bar G}\left(\Mcal[\mathbb P] - \Delta \mathbb H\right)$
\State $\mathbb M_n \gets \gamma_n\left(\mathbb M_n + G_n(\mathbb U_\mathrm{new} - \mathbb U)\right)$
\State $\mathbb U \gets \mathbb U_\mathrm{new}$
\end{algorithmic}
\label{Proposed algorithm}
\end{algorithm}

Here, we consider a periodic contact setting, with period $L$, so that $\Mcal$ is applied in the Fourier domain~\citep{stanleyFFTBasedMethodRough1997,frerotFourieracceleratedVolumeIntegral2019}. \Cref{algorithm} is implemented in the open-source, high-performance rough contact library Tamaas~\citep{frerotTamaasLibraryElasticplastic2020} (\texttt{MaxwellViscoelastic} solver). \added{The spectrum discretization procedure in appendix \ref{app:ml_discrete} is also implemented in Tamaas.}

All simulation results presented in this work use rough surfaces with a self-affine power-spectrum density:

\begin{equation}
\Phi(|\mathbf{k}|)= \begin{cases}C\left(|\mathbf{k}| / k_l\right)^{-2(1+H_e)} & \text { if } k_l \leqslant|\mathbf{k}| \leqslant k_s ; \\ 0 & \text { otherwise, }\end{cases}
\end{equation}
where $\mathbf k$ is the integer wavenumber, $k_l$ and $k_r$ are long and short cutoff wavenumbers respectively, $H_e$ is the Hurst exponent. For representativity~\citep{yastrebovContactRepresentativeRough2012}, we select $k_l = 32$ and $k_s = 256$ and $H_e = 0.8$. Surfaces are discretized with 1024$\times$1024 points and are generated with the algorithm of \citet{wuSimulationRoughSurfaces2000}. Contact areas are computed with a discrete correction~\citep{yastrebovAccurateComputationTrue2017} and averaged over three independent surface samples. A multiplicative factor of 1.15 is applied to simulation area magnitudes to quantitatively match the analytical results\footnote{This is due to the approximate nature of \cref{eq:area}, see \citet{yastrebovInfinitesimalFullContact2015} for an extended discussion.}. The elastic ratio $k = E_\infty / E_0$ was chosen to be $10^{-1}$. Simulation results are marked with circles in all figures, however, only a subset of timesteps is shown for clarity: all time intervals are simulated in 100 geometrically distributed timesteps. Analytical results are shown as full lines.


\section{Results}\label{sec3}

\begin{figure}
    \centering
    \includegraphics{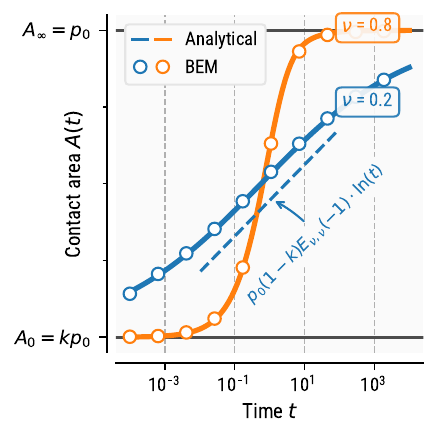}
    \caption{Evolution of the true contact area under constant normal load $p_0$ for a fractional viscoelastic behavior with $\nu = 0.2$ and $\nu = 0.8$ ($k = 0.1$). The exponent $\nu$, which controls the breadth of the relaxation spectrum \added{(the smaller $\nu$, the wider the spectrum)}, greatly influences the time scales over which the contact area transitions from its instantaneous value to its long-time asymptote. Smaller $\nu$ values exhibit the characteristic logarithmic aging behavior over more time-scales, with $\nu = 0.2$ staying in the logarithmic regime for over six decades.}
    \label{monotonic}
\end{figure}

We first simulate the evolution of the contact area under a constant pressure $p(t) = p_0H(t)$, with $H(t)$ the Heaviside function. Applying \cref{eq:area} yields 
\[ A_\mathrm{mono}(t) = p_0J(t).\]
We show in \cref{monotonic} the contact area as a function of time for two values of $\nu$, 0.2 and 0.8. We see the expected asymptotes on short and long time, with a transition happening over a narrow range of time scales for $\nu = 0.8$ \added{(due to its narrow relaxation spectrum)}, whereas $\nu = 0.2$ has a much slower transition, clearly exhibiting a logarithmic evolution regime over about 6 decades. A first order expansion of $A_\mathrm{mono}$ in $\ln(t)$ around $t = 1$ gives the following approximation:
\begin{equation}
    \label{eq:log_A} A_\mathrm{mono}(t) \approx p_0 J(1) + p_0(1 -k)E_{\nu,\nu}(-1)\ln(t).\end{equation}
The multiplicative factor of $\ln(t)$ is therefore proportional to the applied load. This is in agreement with the experimental results of \citet{dillavouNonmonotonicAgingMemory2018}, which have shown proportionality between the applied load and the multiplicative factor of the logarithmic evolution equation for the contact area. The multiplicative factor is also influenced by $\nu$, which additionally controls the range of timescales over which the logarithmic aging regime takes place.

\begin{figure}
    \centering
    \includegraphics{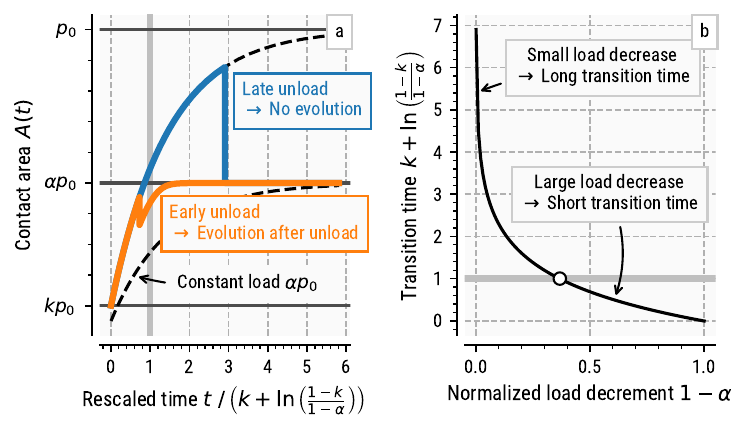}
    \caption{Instantaneous unloading of a single-timescale viscoelastic contact (\added{$\nu =1$}, SLS). (a) shows the evolution of the true contact area when the unloading time $T$ is smaller than the transition time $k + \ln\left(\frac{1-k}{1-\alpha}\right)$: the area keeps evolving after unloading, but does not follow the monotonic curve. When $T$ is larger than the transition time, the area jumps to the asymptotic value for the reduced load $\alpha p_0$, with no evolution thereafter. (b) shows how the transition time depends on the load reduction factor $\alpha$: a larger load reduction has a shorter transition time.}
    \label{transition time}
\end{figure}

In order to study the memory effects of fractional viscoelastic contacts in non-monotonic cases, we next apply a constant normal pressure $p_0$ up to a time $T$, then decrease instantly to a pressure $\alpha p_0$ with $\alpha < 1$. Since the contact area decreases, \cref{eq:t1} is necessary to find the true contact area for times $t > t_m = T$. To illustrate the procedure, we set $\nu = 1$ (Standard Linear Solid) and solve for the unknown time $t_1$:
\begin{align*}
\alpha p_0 & = p_0\int_{-\infty}^{t_1}G(t - \tau)\frac{\mathrm dA}{\mathrm dt}(\tau)\,\mathrm d\tau 
 = p_0 A(0)G(t) + p_0\int_0^{t_1}G(t - \tau)J'(\tau)\,\mathrm d\tau  \\
\Leftrightarrow \alpha & = k + (1-k)e^{-t/k} + \int_0^{t_1} \left(1 + \frac{1-k}{k}e^{-(t-\tau)/k}\right)(1-k)e^{-\tau}\,\mathrm d\tau \\
\Leftrightarrow \frac{1 - \alpha}{1 - k} & = e^{-t_1}\left(1 - e^{-(t - t_1)/k}\right) 
\end{align*}
This last equation can be solved directly when $(t - t_1)/k \gg 1$, we then have $t_1 \approx \ln((1-k)/(1-\alpha))$, which is the time for which $A_\mathrm{mono}(t_1) = \alpha p_0$, the asymptotic contact area for the reduced load. The smaller $\alpha$, the smaller $t_1$ is. Since $t > T$, we therefore know that if:
\[ T \gg k + \ln\left(\frac{1 - k}{1 - \alpha}\right) = T_\mathrm{trans},\]
then the contact area immediately after unloading will be the asymptotic area for a load $\alpha p_0$. This is illustrated by the blue curve in \cref{transition time}a, for which there is no evolution of the contact area after unload: memory of the previous contact area is erased. On the contrary, if the unloading occurs before $T_\mathrm{trans}$, the system will continue to evolve after unloading, creeping up to the asymptotic contact area, as shown by the orange curve in \cref{transition time}a. Note however that it does not follow the monotonic curve with load $\alpha p_0$.

\Cref{transition time}b shows how the transition time $T_\mathrm{trans}$ depends on the load reduction amount $1 - \alpha$. If the load is only slightly reduced, then $T_\mathrm{trans}$ is much longer than the creep timescale (equals to 1 in normalized units), whereas a large load decrease will give a transition time comparable to or smaller than the creep timescale, when $1 - \alpha \geq e^{-1}\approx$~36\% of load reduction.

Fractional viscoelastic materials having a wide relaxation spectrum, their response under non-monotonic loading is not as straightforward as standard linear solids, which have only one relaxation time. To highlight the difference in unloading, we decrease the load linearly with time instead of instantly, i.e.
\[ p(t) = \begin{cases} 0 & t < 0,\\ p_0 & 0 \le t < T, \\ p_0 \left(1 - \frac{t - T}{\Delta T}\right) +\alpha p_0 \frac{t - T}{\Delta T} & T \le t < T + \Delta T, \\ \alpha p_0 & T + \Delta T \le t.\end{cases}\]

\begin{figure}
    \centering
    \includegraphics{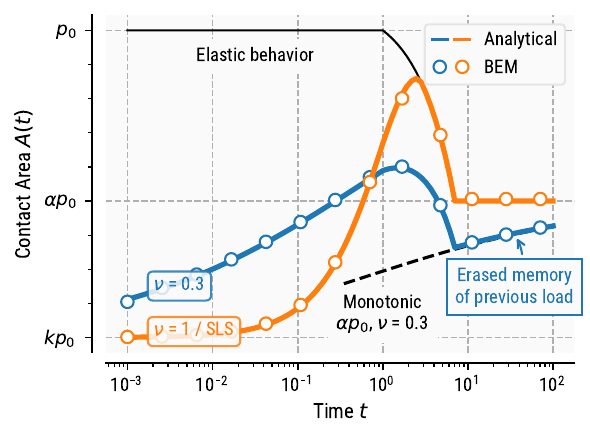}
    \caption{Evolution of the contact area under ramp unloading using the fractional viscoelastic model with $\nu = 0.3$ and $\nu = 1$ (SLS). The black continuous curve shows the purely elastic behavior, which is followed by the SLS curve shortly after the unloading starts at $t = T = 1$. The curve for $\nu = 0.3$ shows a qualitatively similar behavior, but the unloading does not follow the elastic curve. Instead, the contact area decreases down to the monotonic curve with the reduced normal load $\alpha p_0$, and increases with no memory of the previous contact area, in contrast with \cref{transition time}.}
    \label{ramp}
\end{figure}

\Cref{ramp} shows the response for $\nu = 1$ (standard linear solid) and $\nu = 0.3$, with $T = 1$ and $\Delta T = 6$. The full black line shows the elastic behavior (with the long-time modulus). As expected from previous work on viscoelastic contacts under decreasing loads~\citep{greenwoodContactAxisymmetricIndenter2010,papangeloViscoelasticNormalIndentation2021}, the contact area does not decrease as soon as the load decreases, but increases until the time $t_m$, which satisfies the equation:
\[ p_0 J'(t_m) + \frac{p_0(1-\alpha)}{\Delta T}J(t_m - T) = 0.\]
When $t \geq t_m$, we must again solve \cref{eq:t1}, which is expressed as:
\[\frac{p(t)}{p_0} = J(0)G(t) + \int_0^{t_1}G(t - \tau)J'(\tau)\,\mathrm d\tau + H(t_1 - T)\frac{1 - \alpha}{\Delta T}\int_{T}^{t_1}G(t-\tau)J(\tau)\,\mathrm d\tau.\]
For $\nu = 1$, \cref{ramp} clearly shows that soon after $t_m$, the contact area follows the elastic unload curve~\citep{greenwoodContactAxisymmetricIndenter2010} and stops at the asymptotic area for load $\alpha p_0$. However, the model with $\nu = 0.3$ has a different behavior: in the decreasing phase, it does not stop at the asymptote, but instead undershoots it to join with the monotonic curve for load $\alpha p_0$, which contrasts with \cref{transition time}a, where the orange curve does not reach the monotonic loading curve. Memory of the area prior to unloading is thus erased for both the standard linear solid and the fractional viscoelastic linear solid, on \emph{long} time scales after the unload.

\begin{figure}
    \centering
    \includegraphics{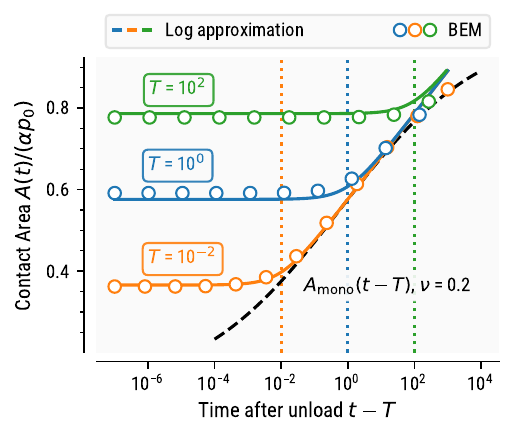}
    \caption{Evolution of the contact area on short timescales after instantaneous load reduction ($\nu = 0.2$), with three unload times $T = 10^{-2}, 10^0, 10^2$, \added{indicated with vertical dotted lines}. The dashed line shows the monotonic evolution under a constant reduced load $\alpha p_0$. On long timescales, all curves join the monotonic curve (shifted in time), but after a time that depends on $T$, the unload time. \added{This is an effect of the change of variable $t - T$. The full curves show the analytical monotonic curves and hightlight that the system follows the monotonic curve even shortly after unload}.}
    \label{dillavou}
\end{figure}

\citet{dillavouNonmonotonicAgingMemory2018} show that memory effects are apparent on time-scales of the order of $T$ after the unload. These time-scales cannot be seen on \cref{ramp}. We show in \cref{dillavou} what happens to the contact area immediately after an instantaneous unload at time $T$, for different values of $T$. Interestingly, the contact area after unload stays constant, contrary to the decreasing regime shown by experiments~\citep{dillavouNonmonotonicAgingMemory2018}. However, as in experiments, there seems to be a clear memory effect: the contact area stays approximately constant for a time comparable to $T$, before joining the monotonic loading curve shifted by $T$. However, this ``memory effect'' can be entirely explained by the change of variables operated in \cref{dillavou}. We have shown in \cref{ramp} that after unloading, the contact area follows the monotonic curve for the reduced load. Let us postulate that this is true for all $t > T$ and $t$ is in the logarithmic regime, i.e.
\begin{equation}\label{eqn:log_approx_memory}\begin{aligned} A(t) & \approx \alpha p_0(J(1) + (1-k)E_{\nu,\nu}(-1)\ln t) \\ & = \alpha p_0\left\{ J(1) + (1-k)E_{\nu,\nu}(-1)\ln T + (1-k)E_{\nu,\nu}(-1)\ln\left(1 + \frac{t - T}{T}\right)\right\} \end{aligned}\end{equation}
The change in contact area after unloading is entirely governed by $\ln(1 + (t-T)/T)$ and $T$ appears to be a memory time over which the area is constant. \Cref{eqn:log_approx_memory} is shown as a continuous line in \cref{dillavou} and perfectly describes the simulation data (circle markers) with no fit parameter, confirming that the contact area \emph{strictly} follows the monotonic curve for the reduced load immediately after unloading, thus erasing all memory of the previous contact state.

\section{Discussion}\label{sec4}

Our results show that a linear fractional viscoelastic framework, which introduces a wide relaxation spectrum, typical of glassy systems~\citep{amirRelaxationsAgingVarious2012}, effectively captures important aspects of the evolution of the contact area under non-monotonic normal loads. In the initial constant-load stage at $p_0$, the contact area exhibits logarithmic aging, with a prefactor controlled by the external load and constitutive parameters which can be independently determined~\citep{huNoteFractionalMaxwell2011}.

Decreasing the external load after a rest time has allowed us to probe the memory effects in the contact area between standard linear solids \added{(i.e. with a single relaxation time)}: this simple system exhibits a behavior change when the rest time is long enough that the contact area is larger than the long-time asymptote for the decreased load. When the load is decreased, the system eventually follows the elastic unloading curve, thus erasing any memory prior to unloading. This means that overloading, followed by unloading, allows the contact area to reach its asymptote much faster than under a constant load. This behavior however depends on the amount of load reduction, which changes the time at which the loading curve crosses the reduced load asymptote.

The same protocol applied to a fractional viscoelastic material yields a different behavior: the contact area during unloading does not follow the elastic curve. It is instead offset: it crosses the reduced load asymptote and joins with the monotonic curve for the reduced load. The system thus evolves with no memory of when the unloading has occurred, even on short time-scales after the unloading. This behavior is qualitatively different from the SLS behavior when unloading before the transition time: the latter still increases after unloading but follows a curve different from the monotonic curve (see orange curve in \cref{transition time}), thus retaining memory of when the unloading occurs. However, when the unloading occurs on small timescales compared to the transition time, the system follows a path asymptotically close to the monotonic curve with the reduced load.

This helps explain why the fractional case falls on the monotonic curve after unloading. Viewing the system as a spectrum of degrees of freedom with a wide distribution of relaxation times \citep{amirRelaxationsAgingVarious2012}, when the unloading occurs, there is a shift in the equilibrium position of all degrees of freedom. However, as we have shown with a single relaxation degree of freedom (the SLS case), degrees of freedom with a transition time shorter than the unload time $T$ immediately jump to the new equilibrium, thus preventing the typical non-monotonic aging observed in many disordered systems~\citep{lahiniNonmonotonicAgingMemory2017}, as the remaining non-equilibrated degrees of freedom still evolve in the same direction towards the new equilibrium. This interpretation is similar to the postulate made by \citet{dillavouNonmonotonicAgingMemory2018} that due to the unilateral contact constraints, a number of out-of-equilibrium degrees of freedom do not contribute to the system evolution, as they are out of contact. The fundamental difference is that in order to reproduce non-monotonic aging, this interpretation requires that the degrees of freedom \added{each have their own relaxation time}, be localized in space (\added{each with its own height}, to resolve a contact condition) and independent from each other, \added{exemplified} by the Winkler foundation model proposed in Ref.~\citep{dillavouNonmonotonicAgingMemory2018}, \added{where each (non-linear) spring-dashpot element has its own relaxation time}. However, in a \added{homogeneous} solid, the displacement of a point on the surface is strongly correlated to all other points due to long range elastic forces: this means that the glassy degrees of freedom are entirely non-local, thus removing the possibility of the non-monotonic evolution of the contact area. In fact, this is easily proved for any linear viscoelastic model. For an instantaneous unload, $t_1$, which is a function of $t$, satisfies the implicit equation:
\[ J(0)G(t) + \int_0^{t_1(t)}G(t - \tau)J'(\tau)\,\mathrm d\tau = \alpha \quad \Leftrightarrow \quad F(t_1(t), t) = \alpha.\]
We can therefore express the derivative of $t_1$ with respect to time:
\[ \frac{\mathrm d}{\mathrm dt}F(t_1(t), t) = 0 \quad \Leftrightarrow \quad t_1'(t) = - \frac{\partial F}{\partial t} \cdot \left(\frac{\partial F}{\partial t_1}\right)^{-1}. \]
The partial derivatives of $F$ can be obtained explicitly:
\begin{align*}
    \frac{\partial F}{\partial t}(t_1, t) & = \int_0^{t_1}G'(t - \tau)J'(\tau)\,\mathrm d\tau + J(0)G'(t),\\
    \frac{\partial F}{\partial t_1}(t_1, t) & = J'(t_1)G(t - t_1).
\end{align*}
Since we always have $J > 0$, $J' > 0$, and $G > 0$, $G' < 0$ (cf.~\cref{rem:transform}) for any linear viscoelastic model~\citep{sharpeSpringerHandbookExperimental2008}, we have $t_1'(t) > 0$ after unload. Since $t_1$ is a time during which $A$ increases, we conclude that $A(t)$ is increasing after unload for any linear viscoelastic model\footnote{This is also true for viscoelasticity with aging, with a similar proof.}.

This finding strongly supports the hypothesis of \citet{dillavouNonmonotonicAgingMemory2018} that the local nature of the material response is responsible for the decrease of the contact area, suggesting the existence of local variables. Such variables could arise from the non-linear viscoelastic behavior of the material, possibly by time-stress superposition, which would lead to stress-dependent relaxation times.

\section{Summary \& Conclusions}\label{sec5}

We have laid out a systematic way of investigating rough contacts with linear fractional viscoelastic materials, both with analytical models and simulations. Using our approach, we were able to probe the evolution of the real contact area in non-adhesive rough contacts under non-monotonic normal loads. Having a wide relaxation spectrum, fractional models can reproduce the logarithmic aging of the true contact area observed for a wide class of materials~\citep{dieterichDirectObservationFrictional1994,dillavouNonmonotonicAgingMemory2018}. However, under non-monotonic conditions, we have shown that memory of the previous contact state is erased when the normal load is reduced: the contact area behaves as if it had experienced a reduced normal force from the beginning of the procedure, even on short time-scales after the unloading. This contrasts with the behavior of standard linear solids, which either jump to the long-time elastic asymptote contact area or follow a path different from the monotonic path after unloading.

The fact that memory is erased after unloading of fractional viscoelastic rough contacts proves that a wide relaxation spectrum is not a sufficient condition to observe memory effects in these systems. We also show that no linear viscoelastic material model can predict a decrease of the true contact area after the load is reduced. Both these facts induce that local variables seem necessary to explain the contact area decrease and the long-term memory effects observed in rough PMMA contacts~\citep{dillavouNonmonotonicAgingMemory2018}.

\subsubsection*{Declarations}
\paragraph*{Ethics approval and consent to participate} N/A
\paragraph*{Consent for publication} N/A
\paragraph*{Funding} N/A
\paragraph*{Authors contributions} LF designed the study. ZL and LF wrote the code and ran calculations. LF derived the analytical model equations. All authors discussed and analyzed results. ZL wrote the first manuscript draft. All authors reviewed the manuscript.
\paragraph*{Acknowledgements}
We thank Vladislav Yastrebov for insightful discussions. The library \textsc{Tamaas}~\citep{frerotTamaasLibraryElasticplastic2020} was used for contact mechanics simulations. Simulation workflow was set up in \textsc{Snakemake}~\citep{molderSustainableDataAnalysis2021a} and figures were produced with \textsc{Matplotlib}~\citep{hunterMatplotlib2DGraphics2007}. A \href{https://archive.softwareheritage.org/swh:1:dir:e09ba0c97ab4998a03d9c0c7ec8806c93410b221;origin=https://github.com/khinsen/mittag-leffler;visit=swh:1:snp:e060cd39a8a52dc41c72e962aa9c3df61dc191fe;anchor=swh:1:rev:f7cdf8f6a9348bc113786011b107d925e839aa10}{Python implementation} of the Mittag--Leffler function created by Konrad Hinsen and David Trémouilles was used in analytical models.
\paragraph*{Data availability}
Codes and data for simulations are available on Zenodo~\citep{liSupplementaryCodesData2025}.

\pagebreak

\begin{appendices}

\section{Equivalent Generalized Maxwell Model}\label{app:ml_discrete}
To approximate $G(t) = 1 + \frac{1-k}{k}E_\nu(-t^\nu/k)$ with a Prony series, we apply a Gauss quadrature rule to the integral expression of the Mittag--Leffler function:
\begin{equation}
    \label{eq:ml_integral} E_\nu(-t^\nu) = \frac{\sin\pi\nu}{\pi}\int_0^\infty P_\nu(\xi) e^{-\xi t}\,\mathrm d\xi, \ \text{where } P_\nu(\xi) = \frac{\xi^{\nu - 1}}{1 + 2\xi^\nu\cos\pi\nu + \xi^{2\nu}}. 
\end{equation}
Following \citet{vernierHomogenizationCompositeMaterials2023, papouliaRheologicalRepresentationFractional2010a}, we split the $[0, \infty[$ domain in three intervals:
\[
\mathbb R^+ = [0, e^{\Lambda/\nu}] \cup [e^{\Lambda/\nu}, e^{\Omega/\nu}] \cup [e^{\Omega/\nu}, \infty[,
\]
and approximate \cref{eq:ml_integral} on each interval.

\paragraph*{Lower interval \texorpdfstring{$[0, e^{\Lambda/\nu}]$}{[0, exp(Lambda/nu)]}}
We use the inequality $\xi\leq e^{\Lambda/\nu}$ to estimate:

\[ 1\geq \int_0^{e^{\Lambda/\nu}} P_\nu(\xi)e^{-\xi t}\,\mathrm d\xi \geq \int_0^{e^{\Lambda/\nu}} P_\nu(\xi)e^{-te^{\Lambda/\nu}}\,\mathrm d\xi = e^{-te^{\Lambda/\nu}}\int_0^{e^{\Lambda/\nu}} P_\nu(\xi)\,\mathrm d\xi\]

\paragraph*{Upper interval \texorpdfstring{$[e^{\Omega/\nu}, \infty[$}{[exp(Omega/nu), infinity[]}}
In the same fashion, we can establish the bounds:

\[ \int_{e^{\Omega/\nu}}^\infty P_\nu(\xi)e^{-te^{\Omega/\nu}}\,\mathrm d\xi = e^{-te^{\Omega/\nu}}\int_{e^{\Omega/\nu}}^\infty P_\nu(\xi)\,\mathrm d\xi \geq \int_{e^{\Omega/\nu}}^\infty P_\nu(\xi)e^{-\xi t}\,\mathrm d\xi \geq 0 \]

\paragraph*{Middle interval \texorpdfstring{$[e^{\Lambda/\nu}, e^{\Omega/\nu}]$}{[exp(Lambda/nu), exp(Omega/nu)]}}
Here we first apply a change of variable $\xi = e^{\theta/\nu}$:
\[ \int_{e^{\Lambda/\nu}}^{e^{\Omega/\nu}}P_\nu(\xi)e^{-\xi t}\,\mathrm d\xi = \frac{1}{\nu}\int_{\Lambda}^{\Omega} P_\nu(e^{\theta/\nu})e^{\theta/\nu} e^{-te^{\theta/\nu}}\,\mathrm d\theta.  \]
Then, we split $[\Lambda, \Omega]$ in $N$ sub-intervals on which we apply a 2-point quadrature rule:

\[ \frac{1}{\nu}\int_{\Lambda}^\Omega P_\nu(e^{\theta/\nu})e^{\theta/\nu} e^{-te^{\theta/\nu}}\,\mathrm d\theta \approx \left(\sum_{n = 0}^{N-1} P_\nu(e^{\alpha_n/\nu})e^{\alpha_n/\nu} e^{-te^{\alpha_n/\nu}} + \sum_{n = 0}^{N-1} P_\nu(e^{\beta_n/\nu})e^{\beta_n/\nu} e^{-te^{\beta_n/\nu}}\right)\frac{\Delta}{2\nu},\]
where:
\begin{align*} 
\alpha_n & = \Lambda + \left(2n + 1 - \frac{1}{\sqrt{3}}\right)\frac{\Delta}{2},\\ \beta_n & = \Lambda + \left(2n + 1 + \frac{1}{\sqrt{3}}\right)\frac{\Delta}{2},\\ \Delta & = \frac{\Omega - \Lambda}{N},\quad n = 1,\ldots,N-1,
\end{align*}
We note $\{\gamma_n\} = \{\alpha_n\} \cup\{\beta_n\}$ the union of quadrature points.

\paragraph*{Remaining integral terms}
We still need to compute the integral term in the bounds we derived, i.e.
\begin{align*}
\int^x P_\nu(\xi)\,\mathrm d\xi & = \int^x \frac{\xi^{\nu - 1}}{1 + 2\xi^\nu\cos\pi\nu + \xi^{2\nu}}\,\mathrm d\xi \\
& = \frac{1}{\nu}\int^{x^\nu} \frac{1}{1 + 2\eta \cos\pi \nu + \eta^2}\,\mathrm d\eta \\
& = \frac{1}{\nu\sin\pi\nu}\arctan\left(\frac{x^\nu - 1}{x^\nu + 1}\tan(\pi\nu/2)\right) +C
\end{align*}
We then have the following definite integrals:
\begin{align*} \int_0^{e^{\Lambda/\nu}}P_\nu(\xi)\,\mathrm d\xi & = \frac{1}{\nu\sin\pi\nu}\left(\arctan\left(\tanh(\Lambda/2)\tan(\pi\nu/2)\right) + \frac{\pi\nu}{2} \right) \\
\int_{e^{\Omega/\nu}}^{\infty}P_\nu(\xi)\,\mathrm d\xi & = \frac{1}{\nu\sin\pi\nu}\left(\frac{\pi\nu}{2} - \arctan\left(\tanh(\Omega/2)\tan(\pi\nu/2)\right)\right)
\end{align*}

\paragraph*{Approximation of $G(t)$}
We choose the lower-bound approximation for both intervals $[0, e^{\Lambda/\nu}]$ and $[e^{\Omega/\nu}, \infty[$, so that we have the following approximation:
\begin{align*}
G(t) & = 1 + \frac{1 - k}{k}E_\nu(-t^\nu / k) \\
& \approx 1 + \frac{1-k}{\pi k}\Bigg(\left(\frac{\pi}{2} + \frac{1}{\nu}\arctan(\tanh(\Lambda/2)\tan(\pi\nu/2))\right)
    e^{-(t/k^{1/\nu})e^{\Lambda/\nu}} \\
&\quad\quad\quad\quad\quad\quad+ \frac{\Delta\sin\pi\nu}{2\nu} \sum_{n} P_\nu(e^{\gamma_n/\nu}) e^{\gamma_n/\nu} 
    e^{-(t / k^{1/\nu}) e^{\gamma_n/\nu}} \Bigg)\\
    & = 1 + \sum_{n}G_ne^{-t/\tau_n}
\end{align*}
This approximation keeps the long-term stiffness $G(\infty) = 1$ of the fractional Zener model.
The sum is then used to characterize the relaxation times and stiffness constants of each Maxwell branch:

\[
\begin{array}{cc}
\text { Relaxation time } \tau_n & \text { Stiffness } G_n \\
\hline k^{\frac{1}{\nu}} e^{-\Lambda / \nu} & \frac{1-k}{\pi k}\left(\frac{\pi}{2}+\frac{1}{\nu} \arctan (\tanh (\Lambda / 2) \tan (\pi \nu / 2))\right) \\
k^{\frac{1}{\nu}} e^{-\gamma_n / \nu} & \frac{1-k}{\pi k} \cdot \frac{\Delta \sin \pi \nu}{2 \nu} P_\nu\left(e^{\gamma_n / \nu}\right) e^{\gamma_n / \nu}
\end{array}
\]
\vspace{0.25cm}

In this work, we chose the numerical values $\Lambda = -7$, $\Omega = 5$, $N = 30$, which gives a total of 61 branches (including the elastic branch). This high number is necessary when $\nu$ is close to 1, as the continuous spectrum approaches a Dirac distribution.
For more details, the reader is invited to look at the notebook \texttt{ml\_approximation.marimo.py} and the file \texttt{python/fractional.py} in the supplementary material~\citep{liSupplementaryCodesData2025}.




\end{appendices}


\bibliography{viscoelastic_rough_contact_area_evolution}

\end{document}